\documentclass[trackchanges,twocolumn]{aastex701}

\usepackage{comment}
\usepackage{booktabs} 
\usepackage{graphicx}  
\usepackage{tabularx}
\usepackage{subcaption}

\usepackage{diagbox}
\usepackage{cancel}
\usepackage[table]{xcolor}
\usepackage{amsmath}

\usepackage{array}

\begin{document}

\title{Jupiter's Evolving Interior as a Source of Episodic Heating on Europa}

\author{Jasmine Sun} 
\affiliation{Department of Physics, Brown University, Providence, RI 02912, USA}
\affiliation{Department of Earth, Environmental and
Planetary Sciences, Brown University,
Providence, RI 02912, USA}
\email[show]{jasmine\_l\_sun@brown.edu}

\author{Benjamin Idini}
\affiliation{Department of Astronomy \& Astrophysics, University of California, Santa Cruz, Santa Cruz, CA 95064, USA}
\email{bidini@ucsc.edu}

\author{Ruth Murray-Clay}
\affiliation{Department of Astronomy \& Astrophysics, University of California, Santa Cruz, Santa Cruz, CA 95064, USA}
\email{rmc@ucsc.edu}

\begin{abstract}

Jupiter's evolving interior can affect the orbital evolution of its moons through frequency-dependent tidal dissipation. We investigate how encounters between the Galilean satellites and Jupiter's oscillation modes affect the stability of the Laplace resonance and Europa's tidal heating. Using the N-body integrator REBOUNDx, we simulate the dynamical orbital evolution of Io, Europa, and Ganymede, with tidal migration timescales derived from three representative models of Jupiter's interior. Encounters with a Jovian tidal resonance can alter the libration state of the Laplace angle, excite orbital eccentricity, and/or destroy the Laplace resonant chain. The outcome of the satellite system depends on which moon experiences enhanced tidal migration. Resonance locks involving Io or Europa can sustain outward migration of the Laplace resonant chain, whereas Ganymede locks destroy the chain. Notably, transient resonance crossings can excite Europa's eccentricity without breaking the Laplace resonance. Under our adopted eccentricity-damping prescription, the baseline tidal power is typically of the order $10^{11}$-$10^{12}$ W, with transient episodes reaching roughly an order of magnitude higher power. These results provide a mechanism for episodic tidal heating on Europa without requiring thermal feedback within the satellites. 

\end{abstract}




\section{Introduction} 

\begin{figure*}
    \centering
    \includegraphics[width=0.9\linewidth]{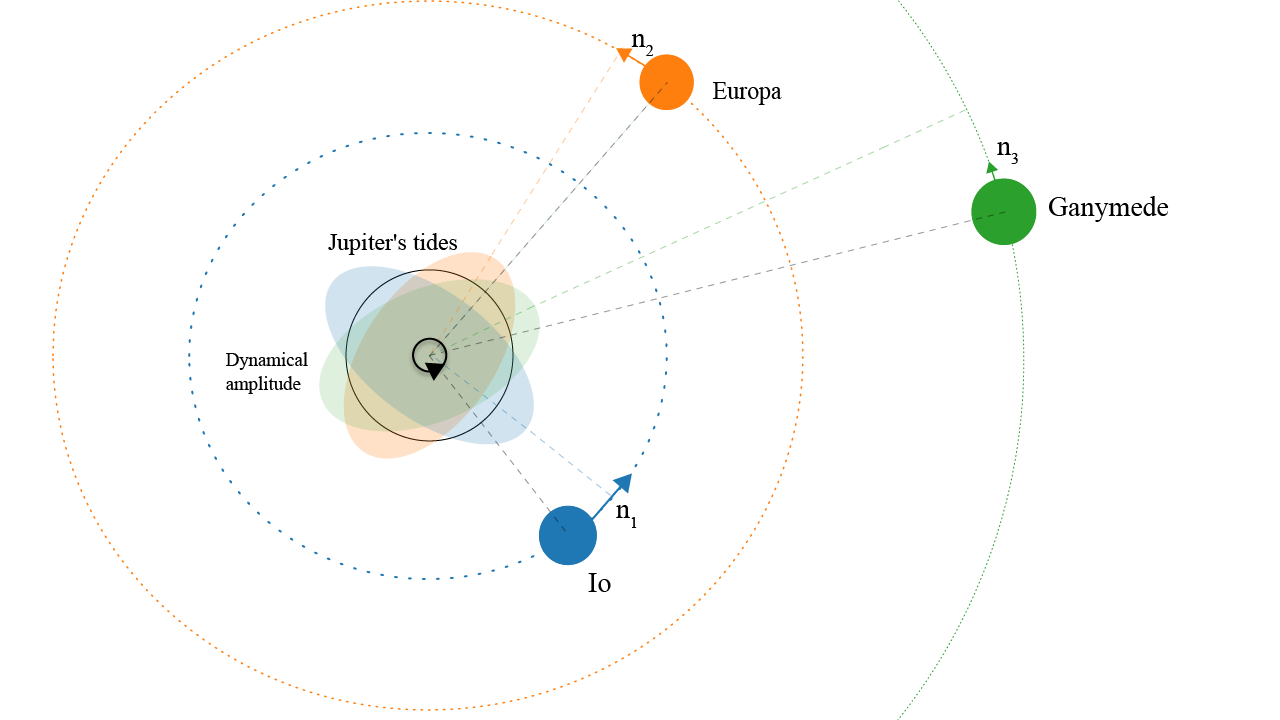}
    \caption{Schematic illustration of tidal forcing by Io, Europa, and Ganymede on Jupiter. Because Jupiter rotates much faster than the moons orbit, its dissipative tidal response is displaced ahead of each moon-Jupiter line, exerting a forward torque that transfers angular momentum from Jupiter's spin to the moons' orbits. The arrows indicate the moons' orbital frequencies \(n_i\), which approximately follow the Laplace commensurability ($n_i/n_{i+1}\approx2$). Each moon forces Jupiter at a different tidal frequency, \(\omega_f=m(\Omega_J-n)\), producing a frequency-dependent dynamical tidal response.}
    \label{fig:concept}
\end{figure*}

Europa hosts a subsurface ocean beneath a tectonically active ice shell, making it a key target in the search for life beyond Earth \citep{vance2023investigating}. Its young ($<100$ Myr) surface preserves a complex record of tectonic evolution, including cross-cutting bands that suggest sequential episodes of deformation \citep{leonard2024global,prockter2002morphology}. Europa's history of deformation may therefore contain clues about how its internal heating has varied over time. 

The Laplace resonance provides the main source of Europa's nonzero eccentricity, sustaining long-term tidal heating that scales approximately as $e_{\rm Eu}^2$. Episodes of enhanced eccentricity are therefore believed to alter the thickness and rheology of Europa's ice shell, with subsequent cooling or refreezing contributing to surface deformation \citep{nimmo2004stresses,howell2018band,kihoulou2025subduction}.

Beyond moon-to-moon interactions, tidal dissipation within Jupiter drives the Galilean satellites outward because Jupiter rotates much faster than the moons orbit. Each moon forces Jupiter at a different tidal frequency $\omega_f$, as illustrated in Fig.~\ref{fig:concept}. Jupiter's tidal response is strongly frequency dependent: the amplitude and phase of the dynamical tide that pushes moons outward vary with $\omega_f$, and both the response and phase can be enhanced when the forcing frequency approaches that of an internal oscillation mode. As Jupiter's interior evolves, its mode spectra changes, causing the associated tidal resonances to sweep through the satellite system and alter the tidal torques acting on the moons. A moon may transiently cross one of these resonances or enter a resonance lock, in which its orbital evolution remains coupled to the evolving mode frequency \citep{fuller2016resonance}. The resulting changes in migration can propagate through the Io--Europa and Europa--Ganymede 2:1 mean-motion resonances, altering the evolution of the full Laplace-resonant chain. 

The interaction of Jupiter's interior with the satellite's orbits produces profound consequences for the dynamics of the system. Beyond the simplest dynamical models of the Laplace resonance, the Laplace angle does not always librate about \(\pi\). In a purely dynamical work, \citet{greenberg1987} demonstrated that stable asymmetric libration centers can emerge when the Galilean satellites occupy a sufficiently deep mean-motion resonance, a result also seen in the primordial migration simulations of \citet{peale2002primordial}. Here, we examine how these three-body mean-motion resonant states respond to changes in satellite eccentricity damping and to differential migration driven by Jupiter's evolving oscillation modes. 

Previous studies have investigated episodic tidal heating driven by feedback between satellite interiors and resonant orbital evolution \citep{yoderpeale1981,ojakangas1986episodic,2004Icar..171..391H,bagheri2026jointdynamical}. Here, we examine a different source of heating variability: changes in the tidal torques exerted by Jupiter as its oscillation modes evolve. This new mechanism does not require a time-dependent satellite rheology and typically produces irregular rather than periodic heating episodes.

After describing our  time-dependent tidal model and numerical methods (\S2–3), we examine asymmetric libration of the Laplace resonance angle in simulations with constant tidal quality factor \(Q\) (\S4), followed by resonance crossings and locks driven by Jupiter's evolving modes (\S5). We then estimate the resulting tidal heating of Europa and discuss its potential tectonic implications (\S6).

\section{Tidal theory}
\label{sec:tidal_theory}

\subsection{Tidal Dissipation Inside Jupiter}
\label{sec:jupiter_tides}

We consider two prescriptions for tidal dissipation within Jupiter. As a baseline, we use a constant tidal quality factor $Q$, for which the semi-major axis migration timescale of a satellite is related to $Q$ by \citep{murray1999solar}
\begin{equation}
    Q = 3 k_2 \frac{m_s}{M_J}
    \left(\frac{R_J}{a}\right)^5 n \tau_a,
    \label{eq:constantQ}
\end{equation}
where $\tau_a=a/\dot a$, $k_2$ is Jupiter's Love number, $m_s$ is the satellite mass, $a$ is its semi-major axis, and $n$ is the satellite orbital frequency.

We also construct a frequency-dependent tidal model in which dissipation is enhanced near resonances between the tidal forcing and Jupiter's internal oscillation modes. For a mode with azimuthal number $m$, the tidal forcing frequency in Jupiter's rotating frame is
\begin{equation}
    \omega_f=m(\Omega-n),
    \label{eq:forcing_frequency}
\end{equation}
where $\Omega$ is Jupiter's spin frequency. When $\omega_f$ approaches a normal-mode frequency $\omega_\alpha$, the dynamical tidal response is resonantly enhanced and dissipation grows, producing narrow regions of rapid orbital migration \citep{fuller2016resonance}.

\begin{figure*}
    \centering
    \includegraphics[width=1\linewidth]{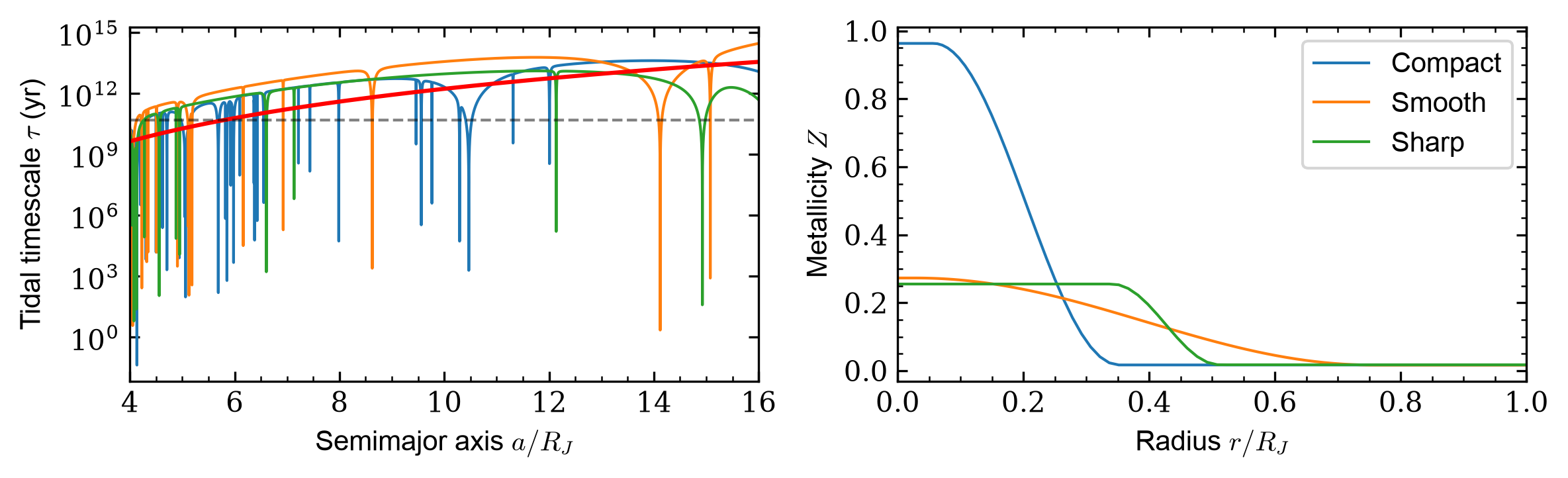}
    \caption{(a) Semi-major axis evolution timescales $\tau_a$ produced by tidal excitation of oscillations in three representative models of Jupiter's dilute core. 
    Sharp features mark tidal resonances between the satellite forcing and Jupiter's oscillation modes for a reference satellite of mass $m_s=10^{25}$~g. The red line shows a constant-$Q$ tidal model for comparison with $Q=10^5$. The dashed line shows the normal-mode evolution timescale $t_{\text{evol}}=50$~Gyr used here. (b) Models of Jupiter's dilute core: Compact \citep{miguel2022jupiter,howard2023jupiter,ziv2024neuralcms}, Smooth \citep{debras2019new}, and Sharp \citep{militzer2022juno,howard2023jupiter,ziv2024neuralcms}. Metallicity is the mass fraction of heavy elements relative to the total mass. All three models fit the same Jupiter average radius.}
    \label{fig:taus}
\end{figure*}

We calculate these frequency-dependent migration profiles for three canonical models of Jupiter's dilute core (Fig.~\ref{fig:taus}b). We use \emph{interiorize} \citep{idini2022gravitational} to construct hydrostatic models of Jupiter and GYRE \citep{townsend2013gyre} to calculate their associated normal modes. Rotation is included through a first-order perturbative correction to the mode frequencies \citep{lai2021jupiter}. We consider tidal forcing of modes with $(\ell,m)=(2,2)$, $(4,4)$, $(4,2)$, $(3,3)$, and $(3,1)$. The detailed mode-frequency corrections, tidal coupling coefficients, and turbulent damping prescription used to construct the migration profiles are given in Appendix~\ref{app:jupiter_tides}.

We approximate the effects of Jupiter's interior evolution by imposing an exponential drift on its normal-mode frequencies, with a common characteristic timescale \(t_{\rm evol}\):

\begin{equation}
    \omega_\alpha(t)
    =
    \omega_\alpha(0)
    \exp\left(\frac{t}{t_{\rm evol}}\right),
    \label{eq:mode_evolution}
\end{equation}
where we adopt $t_{\rm evol}=40$--$50$~Gyr. This value is a fiducial evolutionary timescale motivated by giant-planet resonance-locking models \citep{fuller2016resonance}, rather than a detailed thermal-evolution calculation for Jupiter.

For each satellite, the work done on Jupiter is obtained by summing the energy transfer associated with the modes excited by that satellite. The corresponding orbital migration timescale is
\begin{equation}
    \tau_a
    =
    -\frac{E_{\rm orb}}
    {\sum_\alpha \dot E_\alpha},
    \label{eq:mode_taua}
\end{equation}
where $E_{\rm orb}$ is the satellite's orbital energy and $\dot E_\alpha$ is the energy-transfer rate associated with mode $\alpha$. The resulting $\tau_a(a)$ profiles are shown in Fig.~\ref{fig:taus}a. Between resonances, orbital migration is slow, whereas near individual tidal resonances, $\tau_a$ can decrease by several orders of magnitude. As Jupiter evolves, these resonant features sweep through the satellite system, allowing moons either to cross individual resonances or to become temporarily resonance locked.

\subsection{Heat Production Inside Europa}
\label{sec:europa_heating}

We estimate the tidal heat production inside Europa using the conventional relation \citep{peale1978contribution,peale2003tidally,meyer2007tidal,fuller2016resonance}
\begin{equation}
    \dot{E}
    =
    \frac{\mathcal{G} M_J m e^2}{a\tau_e},
    \label{eqn:edot}
\end{equation}
where $M_J$ and $m$ are the masses of Jupiter and Europa, respectively, $a$ and $e$ are Europa's semi-major axis and eccentricity, and $\tau_e$ is the positive magnitude of its eccentricity-damping timescale. Because the heating rate scales as $e^2$ for fixed $\tau_e$, changes in Europa's eccentricity produced by the orbital dynamics can result in much larger fractional changes in tidal heat production.

\section{Numerical integration of orbital evolution with tides}

We perform simulations using REBOUNDx (version 4.6.1; \citealt{2020MNRAS.491.2885T}), an extension of the N-body integrator REBOUND (version 4.6.0; \citealt{2012A&A...537A.128R}). We include non-conservative orbital evolution using the \texttt{modify\_orbits\_forces} module \citep{2000MNRAS.315..823P,kostov2016tatooine} in REBOUNDx. We prescribe semi-major axis migration and eccentricity damping by the timescales
\begin{equation}
    \tau_a \equiv \frac{a}{\dot a},
    \qquad
    \tau_e \equiv \left|\frac{e}{\dot e}\right|,
\end{equation}
where $\tau_e>0$ denotes the magnitude of the eccentricity-damping
timescale. 

These timescales describe the secular evolution that the nonconservative forces would produce for an isolated satellite. Positive \(\tau_a\) denotes outward migration, while \(\tau_e\) is the positive magnitude of the eccentricity-damping timescale. Because eccentricity damping also modifies the semimajor axis, and resonant interactions exchange angular momentum among the moons, the actual orbital migration rate need not exactly equal \(1/\tau_a\).

To isolate the effects of moon interior rheological feedbacks, we adopt constant eccentricity-damping timescales within each frequency-dependent integration, with \(\tau_{e,\mathrm{Io}}= 6\times10^6\) yr. We set \(\tau_{e,\mathrm{Europa}}=\tau_{e,\mathrm{Ganymede}}=50\tau_{e,\mathrm{Io}}\). These choices are motivated by order-of-magnitude tidal-damping estimates~\citep{murray1999solar}, but are treated as model parameters because satellite dissipation depends on internal structure. In the constant-$Q$ experiments of Section~4, we instead assign the same eccentricity-damping timescale to all three moons and vary its magnitude to examine transitions between libration states (Table~\ref{table:kostov}).

The frequency-dependent tidal profiles are calculated only for the initial normal-mode spectrum. To account for the subsequent evolution of Jupiter's modes without recomputing the oscillation spectrum at every timestep, we map the instantaneous tidal forcing frequency back onto the equivalent location in the initial tidal profile. We assume that the mode frequencies evolve exponentially on a common timescale
$t_{\rm evol}$ (Eq.~\ref{eq:mode_evolution}), exploring values of 40--50 Gyr. These are illustrative interior evolution timescales motivated by the resonance-locking framework of \citet{fuller2016resonance}, rather than directly measured Jovian mode-evolution rates.

For a forcing frequency $\omega_f=m(\Omega_J-n)$,
the corresponding frequency in the initial profile is
\begin{equation}
    \omega_{f,\mathrm{back}}
    =\omega_f \exp\left(-\frac{t}{t_{\rm evol}}\right).
\end{equation}
This gives
\begin{equation}
    n_{\rm back}
    =n\exp\left(-\frac{t}{t_{\rm evol}}\right)
    +\Omega_J\left[1-\exp\left(-\frac{t}{t_{\rm evol}}\right)\right],
\end{equation}
where $n$ is the satellite's instantaneous osculating
mean motion and Jupiter's spin frequency $\Omega_J$
is held fixed. The lookup semimajor axis is
\begin{equation}
    a_{\rm back}
    =\left(\frac{\mathcal{G}M_J}{n_{\rm back}^2}\right)^{1/3}.
\end{equation}

For each satellite $i$, we interpolate the initial tidal profile at its lookup semimajor axis and scale the resulting migration timescale by the satellite mass:
\begin{equation}
    \tau_{a,i}(a_i,t)
    =
    \frac{m_{\rm ref}}{m_i}
    \tau_{a,0}(a_{{\rm back},i}),
    \qquad
    m_{\rm ref}=10^{25}\ {\rm g}.
    \label{eq:mass_scaling}
\end{equation}
Here, $\tau_{a,0}$ is the unaccelerated migration timescale calculated for the reference satellite. The lookup coordinates do not represent the satellite's actual orbital state at an earlier epoch.

The initial semi-major axis of Io is left as a free parameter, with semi-major axis smaller than its current value of $6.03R_J$, since the moons' formation time and migration rate are still uncertain. Initial conditions specific to individual experiments are introduced in each corresponding section. 

We use the WHFast integrator \citep{Rein_2015}, with 15 integration steps per Io orbital period in the main simulations. To reduce computational cost, we accelerate secular evolution by multiplying $\tau_a$, $\tau_e$, and $t_{\rm evol}$ by a common factor $s=10^{-5}$--$10^{-4}$. Such acceleration has been used in previous studies of Galilean-satellite tidal evolution, provided that the dissipative evolution remains sufficiently slow compared with the resonant dynamics \citep{malhotra1991tidal,showman1997tidal, lari2020galilean}.

For frequency-dependent tidal models, we use adaptive tidal-force updates. We identify intervals in the initial tidal profile where the unaccelerated reference migration timescale falls below $10^9$ yr and extend these regions by $0.01R_J$ on either side. When any satellite's lookup coordinate enters one of these regions, we update the tidal migration timescales 10 times per Io orbital period. Outside these regions, the updates follow a coarsely spaced schedule. The tidal forces are held fixed between updates, while the gravitational integration continues at the WHFast timestep.

To test numerical stability near Jovian mode resonances, we repeat selected encounters from saved orbital snapshots with finer tidal updates and weaker secular acceleration. Representative validation integrations use $s=10^{-3}$, 20 WHFast steps per Io orbit, and 10 tidal updates per Io orbit. We compare the resulting resonance-locking behavior, the survival of the Laplace resonance, and eccentricity excitation with the original integrations. Encounters whose qualitative outcomes remain sensitive to the numerical prescription are excluded from our transition classifications.

\section{Jupiter with Constant-Q}

\begin{figure*}
    \centering
    \includegraphics[width=0.92\linewidth]{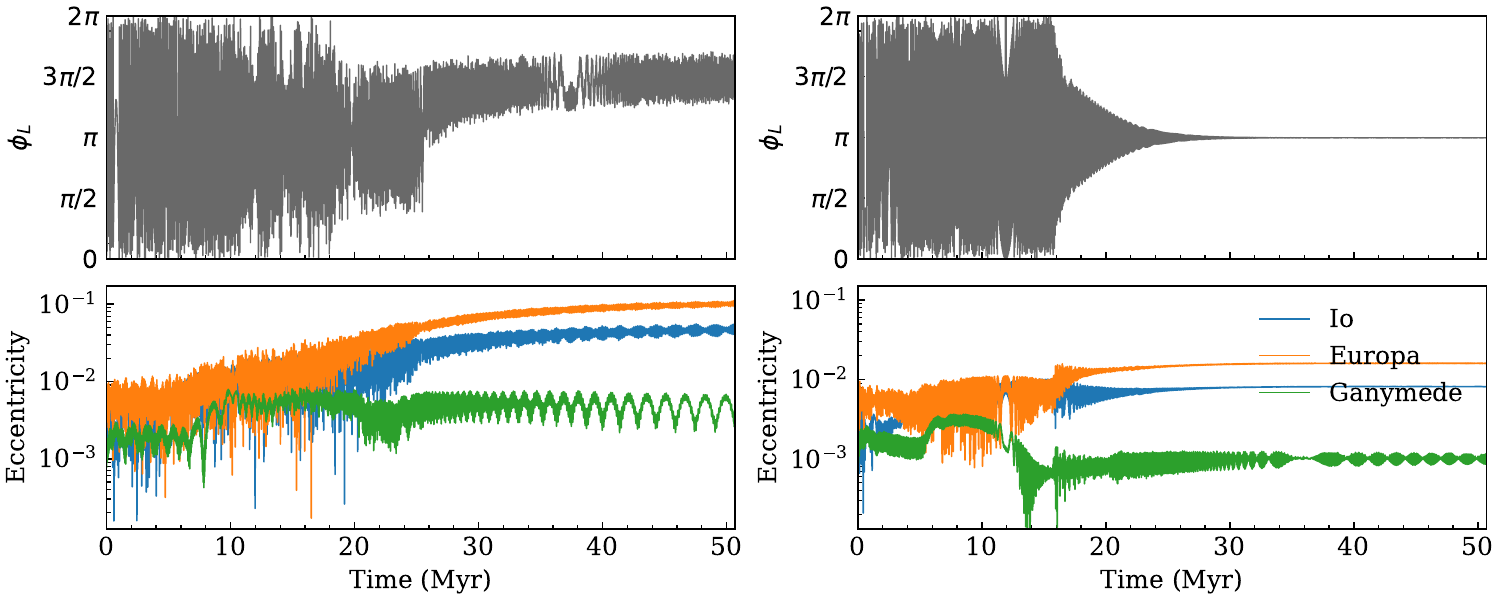}
    \caption{Examples of capture into asymmetric (left) and symmetric (right) libration under constant-\(Q_J=4.4\times10^4\) tidal migration. The satellites begin just wide of the 2:1 mean-motion commensurabilities. Weaker eccentricity damping (left) produces asymmetric libration (\(\tau_{a,\rm Io}/\tau_{e,\rm Io}\sim10\)), whereas stronger damping (right) produces symmetric libration (\(\tau_{a,\rm Io}/\tau_{e,\rm Io}\sim10^3\)).} 
    \label{fig:ALvsSL}
\end{figure*}

\begin{table}
    \centering
    \begin{tabularx}{\columnwidth}{l l l l l }
    \hline
    $\tau_e$ & $\tau_a /\tau_e$ (Io) & \% SL &  \% AL  & $e_{\rm{max}}$ (Eu)  \\
    \hline

$10^9$ yr
& $2$
& 0\%
& 100\%
& 0.30

\\
 
$10^7$ yr
& $2\cdot10^2$
& 20\% 
& 80\%
& 0.15

\\
 
$10^6$ yr 
& $2\cdot10^3$
& 100\%
& 0\%
& 0.01

\\
    \hline 
    \end{tabularx}
    \caption{Outcomes of constant-\(Q_J=4.4\times10^4\) simulations with different prescribed eccentricity-damping timescales, taken to be equal for all three moons. For each parameter set, we run 80 simulations with Io's initial semimajor axis sampled between \(5.4\) and \(5.7R_J\). Percentages give the fraction reaching asymmetric (AL) or symmetric (SL) libration. The ratio \(\tau_a/\tau_e\) characterizes the relative rates of imposed migration and eccentricity damping.}
    \label{table:kostov}
\end{table}

In this section, we use a constant-$Q$ model to examine how eccentricity damping affects capture into the Laplace resonance and the subsequent libration state. We initialize the simulations just wide of the 4:2:1 Laplace resonance, where the ratios between the orbital periods of each pair of moons (Io and Europa, Europa and Ganymede) are 2.02:1. Since moons closer to Jupiter migrate faster due to stronger tidal forces, Io captures Europa and Ganymede in that order into pairwise 2:1 mean-motion resonances. 

\subsection{Capture into Symmetric and Asymmetric Libration}
    \label{sl_and_al}

The Laplace resonance can be understood as a coupled system of two 2:1 mean-motion resonances: an interior resonance between Io and Europa and an exterior resonance between Europa and Ganymede. The respective resonant angles are
\begin{equation}\label{eqn:phi21}
    \phi_{21} = 2\lambda_2-\lambda_1-\varpi_2,
\end{equation}
\begin{equation}\label{eqn:phi32}
    \phi_{32} = 2\lambda_3-\lambda_2-\varpi_2,
\end{equation}
and their difference gives the Laplace angle,
\begin{equation}\label{eqn:phiL}
    \phi_L = \phi_{32}-\phi_{21}
    = 2\lambda_3-3\lambda_2+\lambda_1,
\end{equation}
where $\lambda_i$ and $\varpi_i$ are the mean longitude and longitude of periapsis of satellite $i$ (1: Io, 2: Europa, and 3: Ganymede), respectively.

In the present-day configuration, $\phi_{21}$ and $\phi_{32}$ librate about $\pi$ and $0$, respectively, giving $\phi_L\simeq\pi$. We refer to this as the symmetric configuration. At sufficiently deep resonance, additional stable equilibria can emerge in which the Laplace angle librates about centers displaced from \(\pi\); we refer to these as asymmetric libration states. The stability of asymmetric states has been investigated analytically \citep{yoderpeale1981,greenberg1987}. Beyond the Galilean moons, similar asymmetric equilibria occur in 2:1 resonances between Kuiper Belt objects and Neptune (e.g. \citealt{2005ApJ...619..623M}).  

\begin{figure} 
    \centering
    \begin{subfigure}[t]{0.43\textwidth}
        \includegraphics[width=\textwidth]{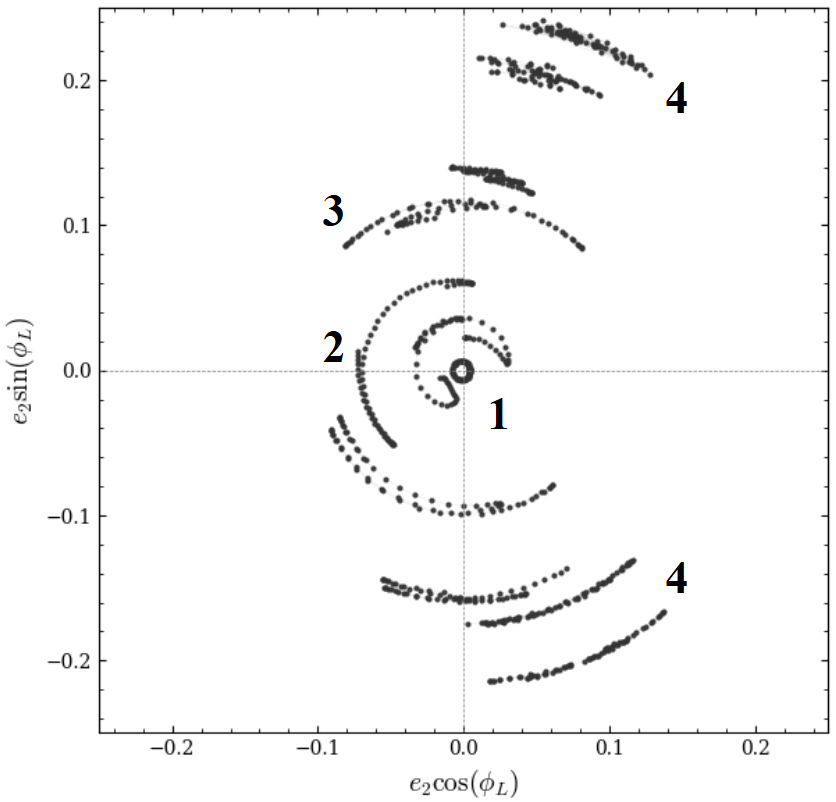}
        \label{fig:contour_w_numbers}
    \end{subfigure}
    \hfill
    \begin{subfigure}[t]{0.56\textwidth}
        \includegraphics[width=\textwidth]{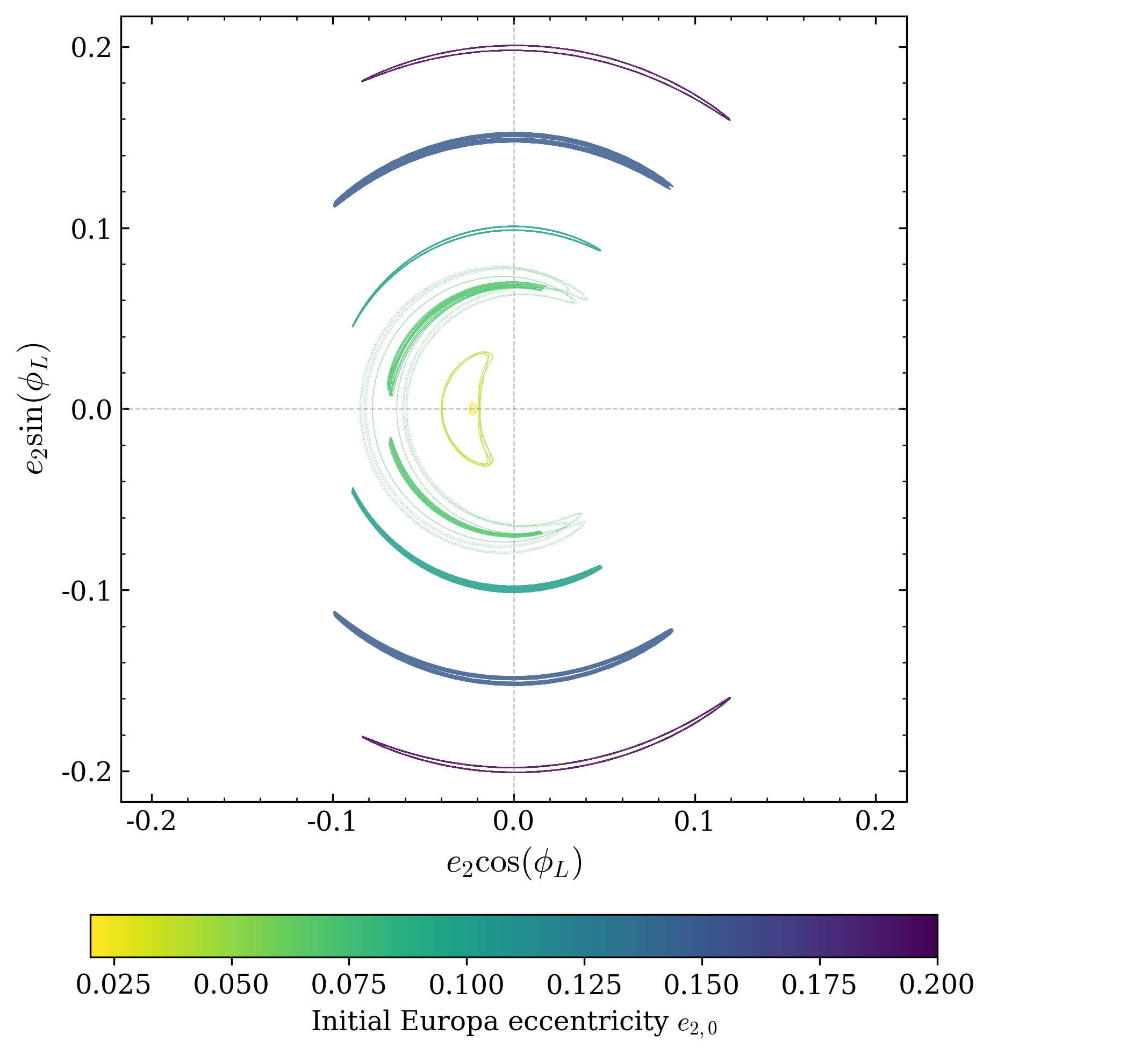}
    \end{subfigure}
        \caption{Projected Laplace phase-space trajectories in \((e_{\rm Eu}\cos\phi_L,e_{\rm Eu}\sin\phi_L)\). (a) Full $N$-body simulation showing circulation (region 1), symmetric libration about \(\phi_L=\pi\) (region 2), and asymmetric libration (regions 3 and 4). (b) Trajectories from simplified simulations placing Io and Ganymede on circular orbits and treating Europa as a massless test particle with \(e_{\rm Eu}=0.02\)–\(0.2\). Successive libration cycles trace nearly closed curves, except near the transition between libration states ($e_2 = 0.07$, green dotted curve). The asymmetric trajectory near $\phi_L = \pi/2$ with $e_2 = 0.1$ uses the initial conditions in Table~\ref{tab:simplified_ICs}.} 
    \label{fig:contours}
\end{figure}

\begin{figure}
    \centering
    \includegraphics[width=0.94\linewidth]{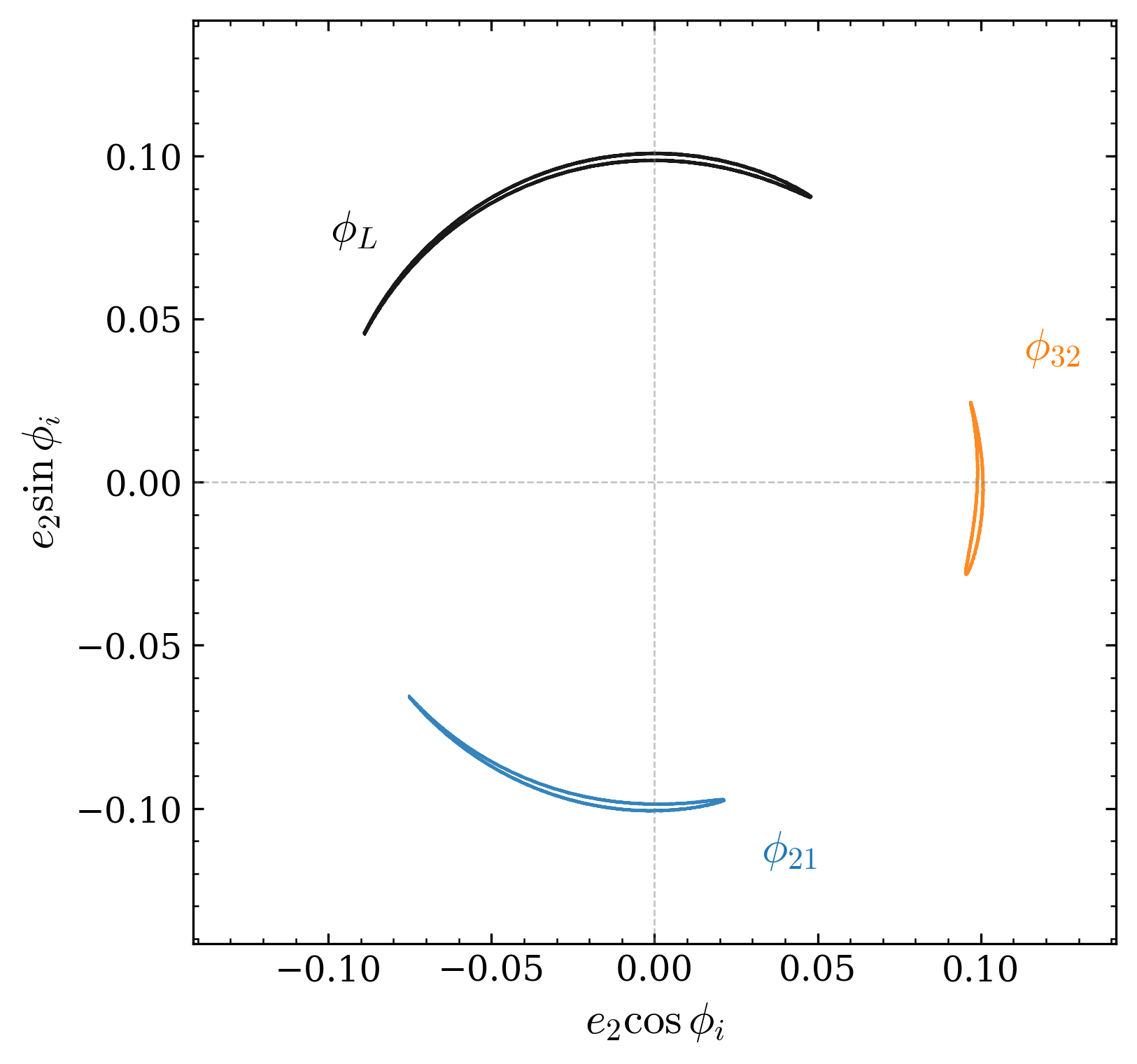}
    \caption{Projected resonant trajectories for the simplified simulation with initial Europa eccentricity $e_2=0.1$, using the initial conditions in Table~\ref{tab:simplified_ICs}. The Io--Europa angle $\phi_{21}$ (blue) librates about a center displaced from $\pi$, while the Europa--Ganymede angle $\phi_{32}$ (orange) remains near $0$. Their difference gives the Laplace angle $\phi_L$ (black), which librates near $\pi/2$. All three trajectories are plotted in $(e_2\cos\phi_i,e_2\sin\phi_i)$.}
    \label{fig:all_resonance_angles}
\end{figure}

\begin{table*}
    \centering
    \begin{tabular}{lcccccc}
        \hline
        Satellite & $m\;(M_\odot)$ & $a$ (AU) & $e$ &
        $\Omega^{a}/\pi$ & $\omega^{b}/\pi$ & $f^{c}/\pi$ \\
        \hline
        
        Io       & $4.49085\times10^{-8}$ & 0.00283131 & 0 &
                    0 & 0.775 & 1.935 \\
        Europa   & 0 & 0.00450000 & 0.1 &
                    0 & 0 & 1.000 \\
        Ganymede & $7.45057\times10^{-8}$ & 0.00713799 & 0 &
                    0 & 1.002 & 2.520 \\
        \hline
    \end{tabular}
    \caption{Initial conditions for the simplified three-satellite configuration used to examine the phase-space structure of the Laplace resonance for $e_{\rm{Eu}}=0.1$. Europa is treated as a massless test particle.
    Semi-major axes are given in AU and masses in solar masses. Angles are given in units of $\pi$. All inclinations are set to zero.
        $ ^{a}$Longitude of the ascending node.
        $ ^{b}$Argument of periapsis.
        $ ^{c}$True anomaly.}
    \label{tab:simplified_ICs}
\end{table*}

In our constant-$Q$ simulations, we find both symmetric and asymmetric libration of $\phi_{L}$. For sufficiently strong eccentricity damping, the Laplace angle evolves toward the present-day symmetric configuration and librates about $\pi$. When eccentricity damping is weaker, the system reaches higher eccentricities and can instead librate about centers displaced from $\pi$, typically near $\sim\pi/2$ or $\sim3\pi/2$ (Fig.~\ref{fig:ALvsSL}). In our simulations, the libration center generally becomes more strongly displaced from $\pi$ as the eccentricity increases. These results are qualitatively consistent with the deep-resonance solutions of \citet{greenberg1987} and with the offset libration centers obtained during the primordial-migration calculations of \citet{peale2002primordial}. In these asymmetric states, the displacement of $\phi_L$ from $\pi$ arises primarily from a shift in the libration center of the inner Io--Europa resonant angle $\phi_{21}$, while $\phi_{32}$ remains closer to its symmetric center.

To investigate the phase-space structure underlying these configurations, we construct projected resonant trajectories using $(e_2\cos\phi_L,e_2\sin\phi_L)$. In the full $N$-body integrations, these trajectories do not form closed curves (Fig.~\ref{fig:contours}a). Similar phase-space representations using individual 2:1 resonant angles have been used to illustrate resonant states in the Galilean satellites \citep{peale2002primordial}. The lack of closed trajectories in the full system may reflect the presence of additional degrees of freedom and indicates that $e_2$ alone does not provide a complete canonical description of the Laplace dynamics.

To clarify the dynamical origin of asymmetric libration, we also examine a simplified configuration in which Europa is massless and Io and Ganymede are initialized on circular orbits.  This arrangement provides insight into the qualitative behavior of the Laplace angle (Equation \ref{eqn:phiL}) because the two-body resonance angles in Equations (\ref{eqn:phi21}) and (\ref{eqn:phi32}) both refer to eccentricity-type mean-motion resonances associated with the eccentricity of the central body (Europa).  The moons' semimajor axes and orbital phases are adjusted from JPL Horizons \citep{Giorgini1996_Horizons, Jacobson2021_JUP365} values to place the system near the 2:1 commensurabilities. Initial conditions for one case with $e_2=0.1$ are given in Table~\ref{tab:simplified_ICs}. In this configuration, several projected trajectories trace nearly closed curves over consecutive libration cycles (Fig.~\ref{fig:contours}b). For the $e_2=0.1$ case, we also plot the two individual 2:1 resonant angles (Fig.~\ref{fig:all_resonance_angles}). The inner angle $\phi_{21}$ librates near $3\pi/2$, while $\phi_{32}$ remains near $0$, giving asymmetric libration of $\phi_L$ near $\pi/2$. 

The appearance of closed trajectories in the simplified system is consistent with an approximately one-degree-of-freedom description of the dominant resonant motion. However, $e_2$ is not the canonical conjugate momentum to $\phi_L$: in the averaged Hamiltonian, the corresponding action is related to Europa's orbital angular momentum, such as $\Phi_4$ defined in \citet{malhotra1991tidal}, rather than $e_2$ alone. Thus, $(e_2\cos\phi_L,e_2\sin\phi_L)$ should be regarded as a useful visualization of the simplified resonant dynamics rather than an exact canonical phase space. A more complete Hamiltonian treatment remains an area for future work. 

\begin{figure}
    \centering
    \includegraphics[width=.97\linewidth]{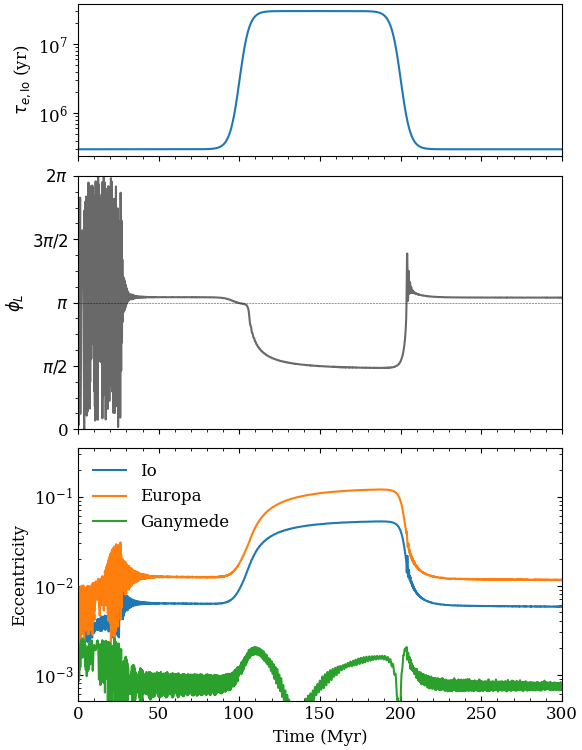} 
    \caption{Evolution under a temporary reduction in Io’s eccentricity damping using the Constant-$Q$ Jupiter model. Top: imposed smooth top-hat variation in \(\tau_{e,\mathrm{Io}}\), increasing from \(3\times10^5\) to \(3\times10^7\) yr during the middle third of the integration before returning to its initial value. Middle: corresponding evolution of the Laplace angle \(\phi_L\). Bottom: satellite eccentricities. The imposed change in damping is accompanied by a transition from symmetric to asymmetric libration while keeping the Laplace resonance intact. Symmetric libration resumes when stronger damping is restored.}
    \label{fig:311_consq_episode}
\end{figure}

\subsection{Transitions Between Libration States}
    \label{subsection:edamp}

Having found that weaker eccentricity damping favors asymmetric libration during capture, we next examine whether a temporary change in damping can alter the libration state of an established Laplace resonance. We temporarily increase Io's eccentricity-damping timescale, $\tau_e$, from $3\times10^5$ to $3\times10^7$ yr during the middle third of the integration, before returning it to its initial value. The variation in $\tau_e$ is implemented as a smoothed top-hat in \(\log \tau_e\). This prescription is not intended to represent a specific thermal cycle in Io.

Figure~\ref{fig:311_consq_episode} confirms that the Laplace angle transitions from symmetric to asymmetric libration as damping weakens, then returns to symmetric libration when stronger damping is restored. The Laplace resonance remains intact throughout these transitions.

\section{Jupiter with Frequency-Dependent $Q(\omega)$}

\begin{figure*}
    \centering
    \includegraphics[width=0.99\linewidth]{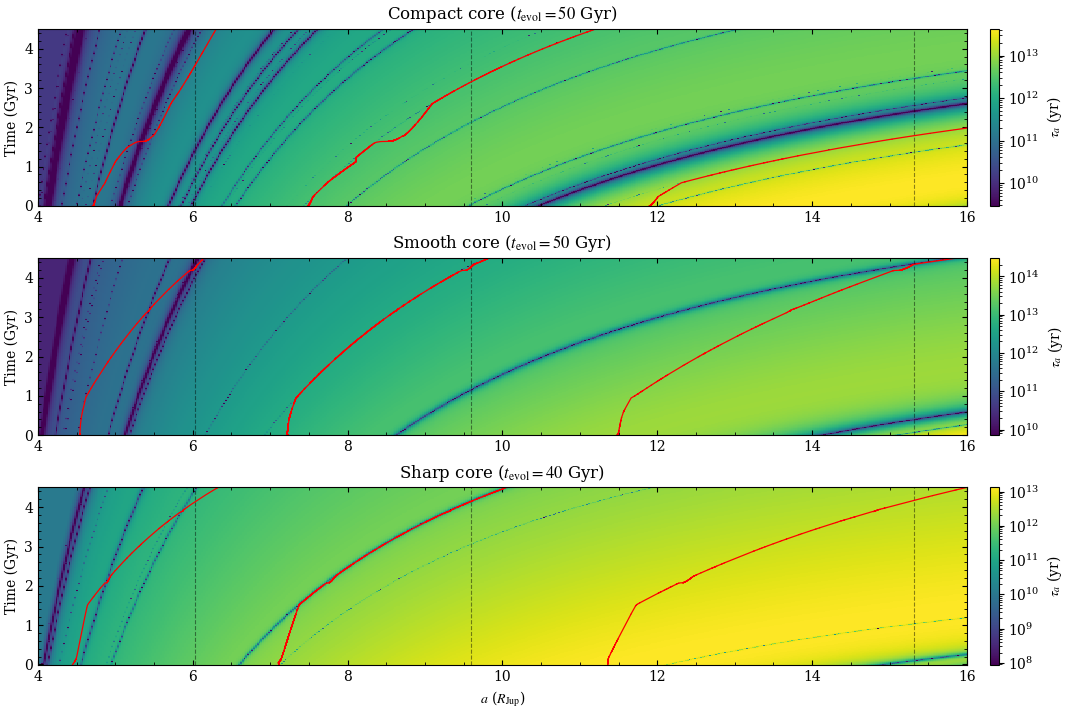}
    \caption{Evolution of Jupiter's tidal migration profiles for the Compact (top), Smooth (middle), and Sharp (bottom) interior models. Colors indicate the prescribed migration timescale \(\tau_a\) as a function of semimajor axis and time; narrow minima trace the outward drift of Jovian tidal resonances. Red curves show the semimajor axes of Io, Europa, and Ganymede in representative simulations. The adopted mode-evolution timescales are \(t_{\rm evol}=50\), 50, and 40 Gyr, respectively. Vertical dashed lines mark the moons' present-day orbital distances. The orbital elements of these three simulations are in Figures~\ref{fig:231_core0_params} and~\ref{fig:smooth_sharp}.}
    \label{fig:231_328_236.png}
\end{figure*}

\begin{figure}
    \centering
\includegraphics[width=.93\linewidth]{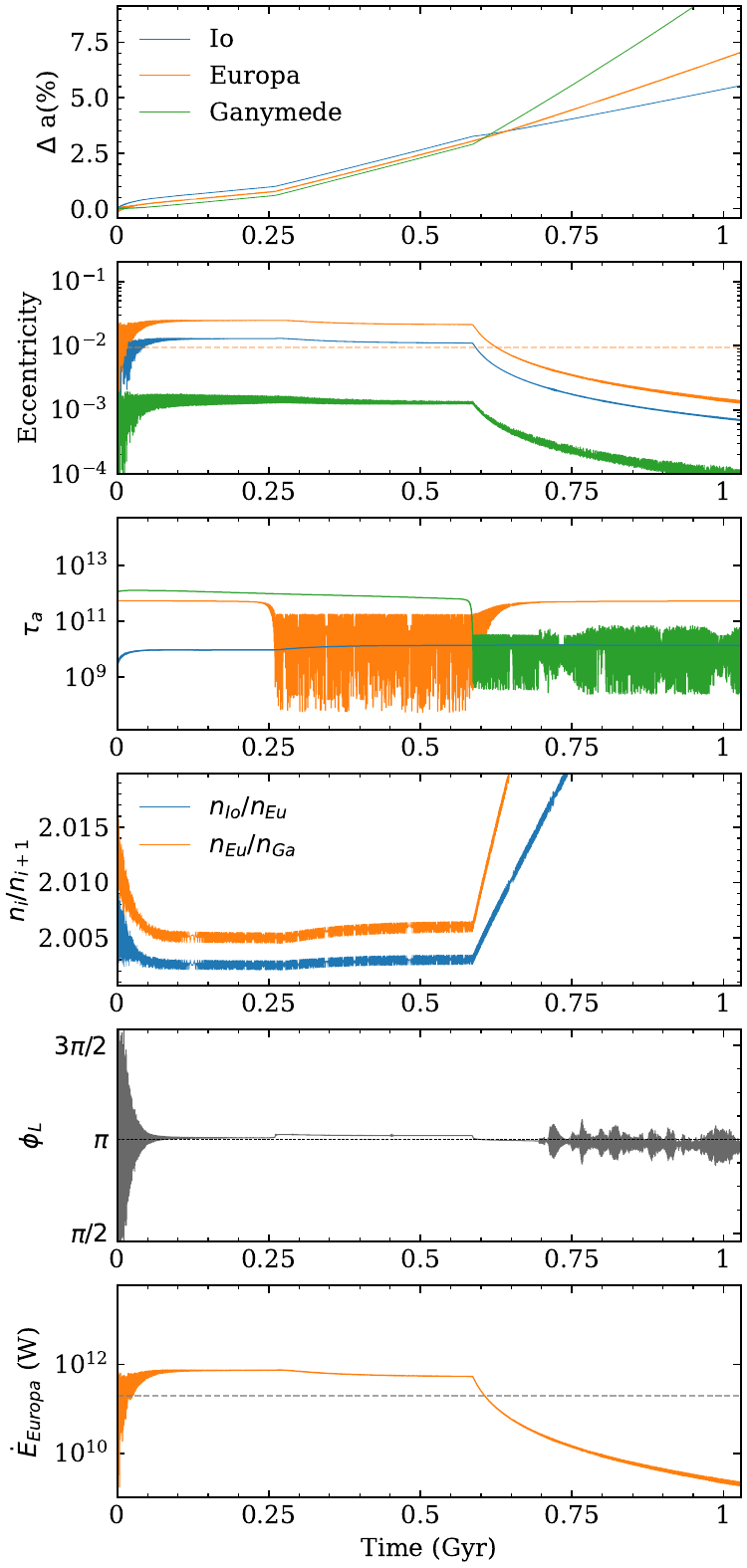}
    \caption{First 1 Gyr of orbital evolution in the Compact core simulation (Fig.~\ref{fig:231_328_236.png},top). The satellites begin just wide of the 2:1 commensurabilities, with initial adjacent mean-motion ratios of 2.01. Europa becomes locked to a Jovian tidal resonance at approximately 0.25 Gyr. Ganymede subsequently locks near 0.6 Gyr, disrupting the Laplace resonance.}
    \label{fig:231_core0_params}
\end{figure}

\begin{figure*}[t]
    \centering
    \begin{minipage}[t]{0.44\textwidth}
        \centering
        \includegraphics[width=\linewidth]{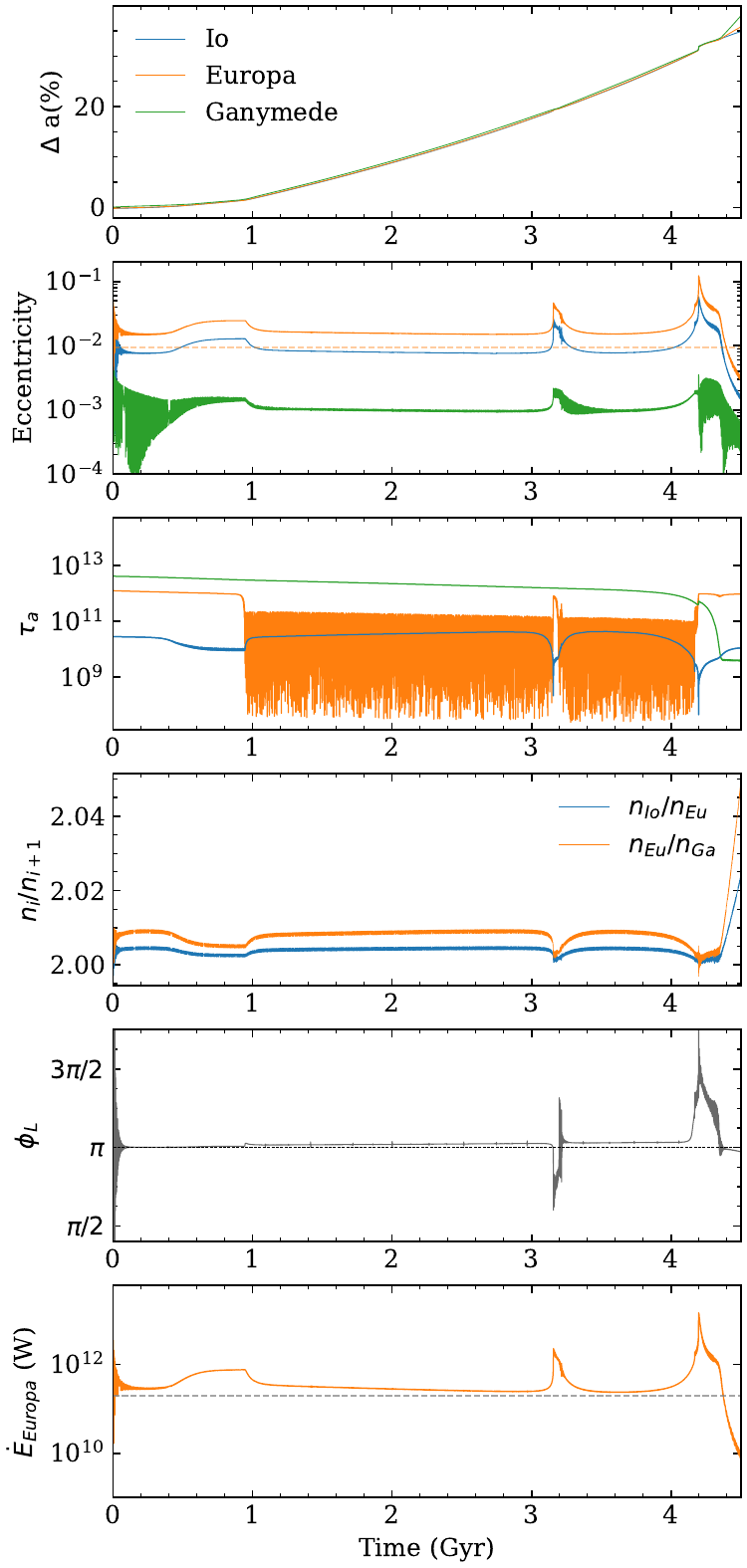}
        \vspace{0.5em}
        \text{(a) Smooth core}
    \end{minipage}
    \hspace{1cm}
    \begin{minipage}[t]{0.44\textwidth}
        \centering
        \includegraphics[width=\linewidth]{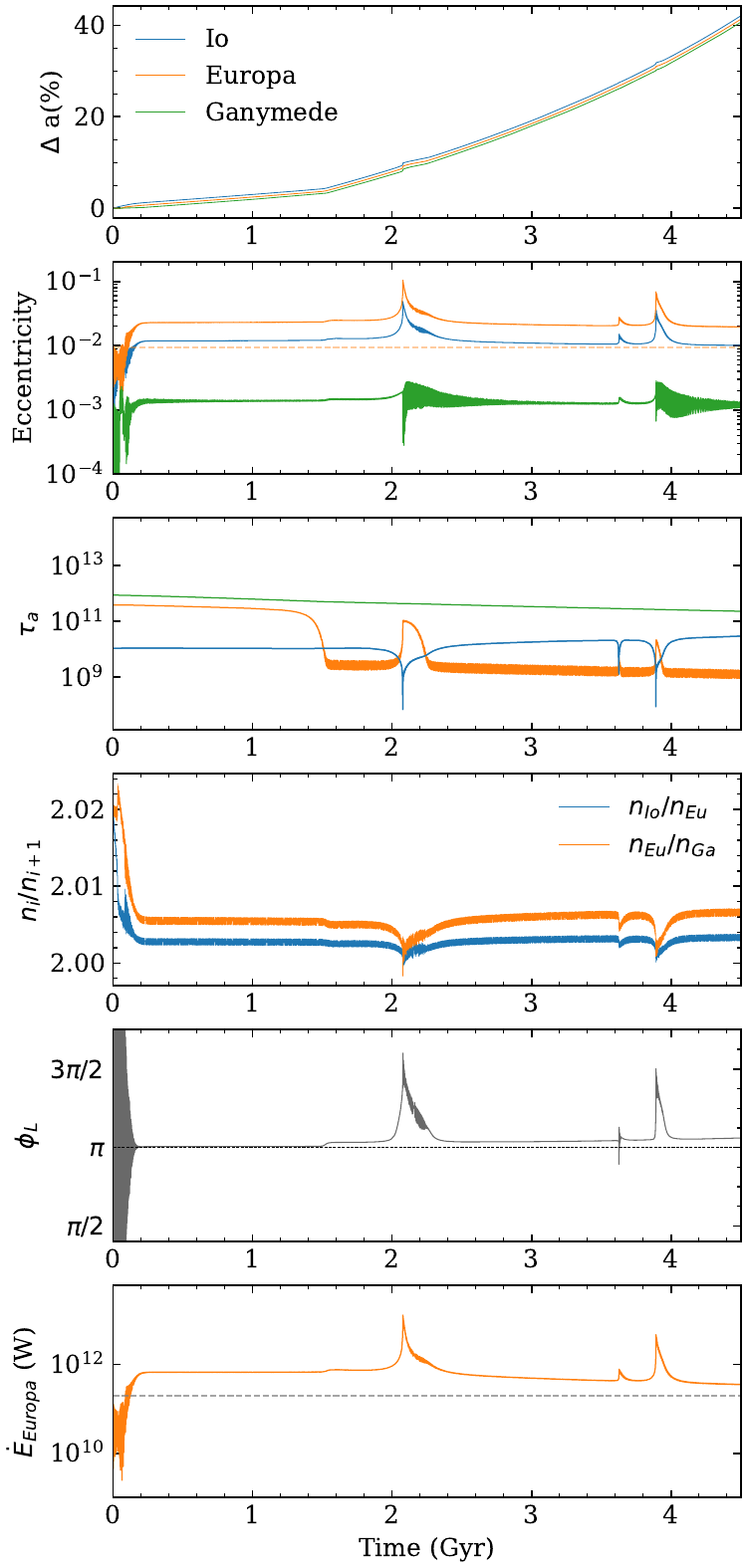}
        \vspace{0.5em}
        \text{(b) Sharp core}
    \end{minipage}

    \caption{Orbital evolution in representative Smooth-core (a) and Sharp-core (b) simulations. (a) Starting from exact 2:1 period ratios, Io locks to a Jovian tidal resonance near 0.4 Gyr, followed by Europa near 1 Gyr. Subsequent Io resonance crossings excite the satellite eccentricities before a Ganymede lock disrupts the Laplace resonance at approximately 4.3 Gyr. 
    (b) Starting from period ratios of 2.02, the satellites are captured into the Laplace resonance through convergent migration. Europa locks near 0.8 Gyr, while later Io resonance crossings produce temporary eccentricity enhancements without disrupting the Laplace resonance. The rapid variations in Europa's prescribed \(\tau_a\) in panel (a) likely reflect the steep gradient of the narrow Jovian tidal resonance to which Europa is locked, as opposed to the wider resonance Europa is locked to in (b).}
    \label{fig:smooth_sharp}
\end{figure*}

We now investigate how Jupiter's frequency-dependent tidal response affects the coupled orbital evolution of Io, Europa, and Ganymede. As Jovian mode resonances sweep through the satellite system, encounters can alter the migration of individual moons, excite eccentricity, shift the Laplace angle's libration center or disrupt the resonance. We hold the satellite eccentricity-damping timescales fixed within each integration and examine the effects of strongly varying tidal migration rates.

In models with frequency-dependent $Q(\omega)$, Jupiter's oscillation modes produce narrow regions of enhanced tidal dissipation, which appear as sharp minima in $\tau_a$ (Fig.~\ref{fig:taus}). As Jupiter's interior evolves, the mode frequencies and their associated interior--orbital resonances sweep through the satellite system. Away from these resonances, tidal migration is substantially slower than in our constant-$Q$ model, whereas near resonance the migration rate can be strongly enhanced. Consequently, convergent migration driven solely by the background tidal dissipation would require substantially longer than in the constant-$Q$ simulations, by approximately two orders of magnitude for the models considered here. We initialize the positions of the satellites just wide of $2:1$ pairwise orbit ratios.

\subsection{Jupiter Models}

We study orbital interactions with three canonical models of Jupiter's interior that roughly represent core geometries previously suggested by different groups based on interpretations of Juno's gravity data (Fig.~\ref{fig:taus}). The initial locations of interior-orbital resonances vary strongly across different core structures, driven mostly by the thickness and radial position of the stratified cavity that hosts the planet's g-modes. When out of resonance, the migration timescale of moons is much slower than that of the constant-$Q\sim 10^5$ model and scales proportionally to the volume of convective fluid inside the planet.

The compact-core model is the closest of our three models to the traditional picture of a gas giant in which most heavy elements are concentrated toward the planet's center. The model remains fluid throughout, but contains a deeply settled stably stratified layer and therefore captures seismological features similar to proposed Jupiter models with relatively compact dilute cores
\citep{miguel2022jupiter,howard2023jupiter,ziv2024neuralcms}.

The origin of Jupiter's heavy-element gradient remains uncertain. Redistribution of material from an initially compact core may occur through dissolution and erosion of heavy elements into metallic hydrogen \mbox{\citep{wilson2012rocky,wilson2012water}}, while the deposition and settling of accreted planetesimals during formation can also establish primordial compositional gradients \citep{lozovsky2017jupiter}.

The other two models (Fig.~\ref{fig:taus}), smooth \citep{debras2019new} and sharp \citep{militzer2022juno}, constitute extended dilute core structures proposed by independent analyses of the same Juno observational constraints. These models are geophysically consistent, but they remain in tension with reproducing and preserving the outward extending stratified layers without mixing the material over Jupiter's lifetime \citep{vazan2018jupiter,muller2020challenge}.

\begin{table*}[htbp]
\centering
\renewcommand{\arraystretch}{1.3}
\setlength{\tabcolsep}{6pt}

\begin{tabular}{lcccc}
\hline

\scriptsize\diagbox[width=2.4cm,innerleftsep=-1.2cm,innerrightsep=-0.9cm]{Current state}{Next resonance\\ \hspace{15pt} encounter}
& None & Io & Eu & Ga \\ \hline

\parbox[t]{3.05cm}{\raggedright No resonance locking} &
\parbox[t]{3.05cm}{\raggedright LR maintained} &
\parbox[t]{3.05cm}{\raggedright Io locks} &
\parbox[t]{3.05cm}{\raggedright Europa locks, LR may break.} &
\parbox[t]{3.05cm}{\raggedright Ganymede locks; LR breaks}
\\ \hline

\parbox[t]{3.05cm}{\raggedright Io locked} &
\parbox[t]{3.05cm}{\raggedright LR maintained} &
\parbox[t]{3.05cm}{\raggedright ---} &
\parbox[t]{3.05cm}{\raggedright Europa locks, replacing Io's lock; LR survives or breaks.} &
\parbox[t]{3.05cm}{\raggedright Ganymede locks; LR breaks}
\\ \hline

\parbox[t]{3.05cm}{\raggedright Europa locked} &
\parbox[t]{3.05cm}{\raggedright LR maintained, accelerated migration, or Europa carries Ganymede away, breaking LR.} &
\parbox[t]{3.05cm}{\raggedright Eccentricity excitation; Europa remains locked; LR survives or breaks. If Io locks, LR may break.} &
\parbox[t]{3.05cm}{\raggedright ---} &
\parbox[t]{3.05cm}{\raggedright Ganymede locks, LR breaks}
\\ \hline

\parbox[t]{3.05cm}{\raggedright Ganymede locked} &
\parbox[t]{3.05cm}{\raggedright LR remains broken} &
\parbox[t]{3.05cm}{\raggedright LR remains broken} &
\parbox[t]{3.05cm}{\raggedright LR remains broken} &
\parbox[t]{3.05cm}{\raggedright ---}
\\ \hline

\end{tabular}

\caption{Qualitative outcomes of encounters with evolving Jovian tidal resonances for systems initially in Laplace resonance (LR). Rows indicate the satellite initially locked to a Jovian mode, and columns indicate the satellite encountering a Jovian tidal resonance next. The outcomes describe the behavior found in our numerical experiments and are not necessarily exhaustive.}
\label{table:outcomes_old}

\end{table*}

\subsection{Interactions with Jupiter's Interior}

We integrated the orbital evolution of the Galilean satellites using the tidal migration timescales calculated for the three Jupiter interior models. In each simulation, Io was initially placed between $4.5$ and $5.5\,R_{J}$, and the satellites were initialized in the Laplace resonance. The initial orbital phases were chosen so that Io--Europa conjunctions occur near Europa's apocenter, and Europa--Ganymede conjunctions occur near Europa's pericenter.

The different outcomes resulting from the moons interacting with Jupiter's evolving oscillation modes are summarized in Table~\ref{table:outcomes_old}. These outcomes were identified by manual inspection of a broad set of initial conditions, tidal parameters, and interior models of Jupiter. They are organized based on which satellite encounters a sweeping tidal resonance and whether a satellite is currently locked. Representative examples for each of the Jupiter models containing the main dynamical behaviors are shown in Fig.~\ref{fig:231_328_236.png}. The corresponding orbital elements are shown in Figs.~\ref{fig:231_core0_params} and~\ref{fig:smooth_sharp}a,b.

The outcome of a Jovian tidal resonance encounter depends on which satellite experiences the enhanced torque and whether the three moons remain coupled through the Laplace resonance. When Io locks to a Jovian tidal resonance, its accelerated outward migration can carry Europa and Ganymede with it while preserving the resonant chain. Europa locking can also transport all three moons outward, provided the coupled resonances remain intact. In other cases, Europa and Ganymede migrate outward together while Io is left behind, disrupting the Laplace resonance. Ganymede locking does not carry the chain outward in our simulations, as Ganymede migrates away from the inner two moons, breaking the Laplace resonance.

Successive encounters can also change which satellite remains locked to a Jovian mode. For example, in the Smooth-core simulation (Fig.~\ref{fig:smooth_sharp}), Io is initially resonance locked, but Europa's subsequent capture into another Jovian mode resonance displaces Io from its previous lock. The migration of the resonant chain then follows the new locking configuration. These transitions can occur without the initially locked moon crossing the center of its Jovian tidal resonance, since changes in the coupled orbital migration can move it away from the stable locking location. 

\subsubsection{Resonance Locking}

If a tidal resonance associated with a Jovian mode sweeps outward faster than a moon would migrate under the background tidal dissipation, the moon can be captured into a resonance lock. Except at orbital distances very close to Jupiter ($a\lesssim4.5R_{J}$), migration driven by the non-resonant convective background is relatively slow (Fig.~\ref{fig:taus}a). The satellites therefore migrate slowly outward until a tidal resonance approaches their orbital location.

A resonance lock can be maintained on the flank of a Jovian tidal resonance when a satellite's net outward migration rate matches the drift of the mode resonance \citep{fuller2016resonance}. 

For our prescribed exponential mode evolution, the
resonance location migrates outward at the rate
\begin{equation}
    \frac{\dot a_{\rm res}}{a_{\rm res}}
    =
    \frac{2}{3t_{\rm evol}}
    \left(\frac{\Omega_J}{n_{\rm res}}-1\right),
\end{equation}
corresponding to a migration timescale
\begin{equation}
    \tau_{\rm res}
    =
    \frac{3t_{\rm evol}}{2}
    \frac{n_{\rm res}}{\Omega_J-n_{\rm res}}.
    \label{eq:resonance_drift}
\end{equation}
For an isolated satellite, resonance locking requires its tidal migration rate to match the outward drift of the resonance. In the coupled Galilean system, the relevant condition instead involves the satellite's net migration rate, including angular-momentum exchange through the mean-motion resonances.

If the moon moves toward the resonance center, the stronger tidal dissipation drives migration faster than the resonance itself, pushing the moon outward relative to the resonance. Conversely, if the moon is too far from the resonance center, tidal migration becomes too slow, and the resonance catches up. 

\begin{figure}
    \centering
    \includegraphics[width=0.99\linewidth]{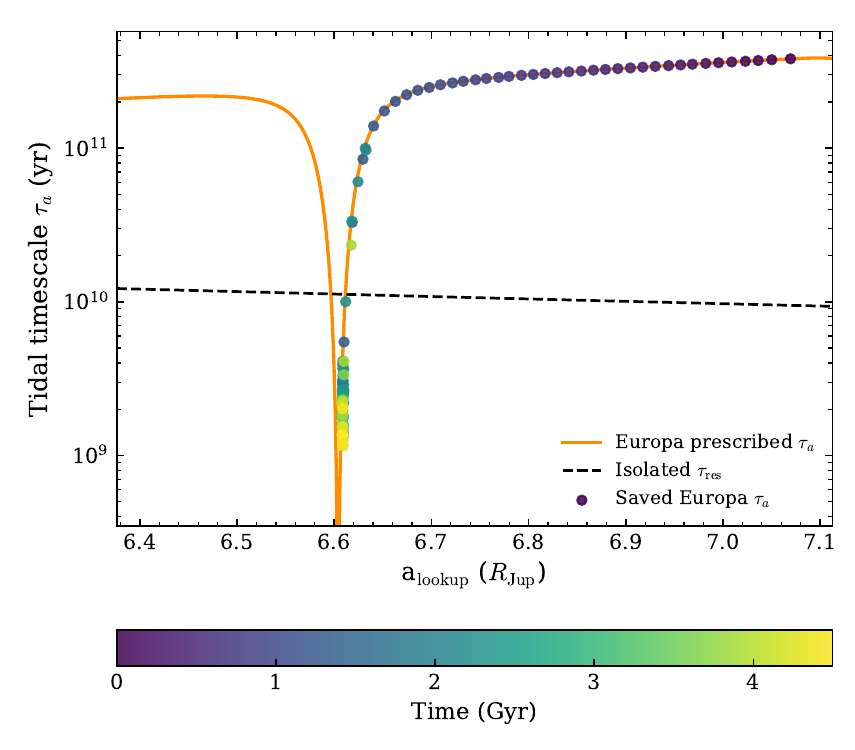}
    \caption{Europa's resonance lock in the Sharp-core simulation, shown relative to Jupiter's initial tidal migration profile, $\tau_{a,0}(a)$. The dots mark Europa's position in the initial profile, evaluated using $a_{\rm lookup}$, as Jupiter's mode frequencies evolve. The horizontal dashed line marks \(\tau_{\rm res}\) (Equation~\ref{eq:resonance_drift}), the tidal migration timescale required for an isolated satellite to remain locked to the evolving Jovian mode. Europa moves toward the resonance minimum and settles on a flank where the prescribed migration timescale is shorter than the isolated-satellite locking timescale. This offset suggests that a stronger tidal torque on Europa is required to sustain the outward migration of the coupled Laplace-resonant chain.} 
    \label{fig:europa_locking_location}
\end{figure}

In the coupled three-moon system, however, a satellite's net migration rate also depends on resonant interactions with the other moons. A change in another satellite's tidal torque can displace a previously locked moon from this equilibrium, even without passage through the center of the Jovian resonance. Figure~\ref{fig:europa_locking_location} illustrates this distinction for a Europa resonance lock. Europa moves toward the stronger-dissipation portion of the Jovian resonance, where its tidal torque can contribute to the migration of the coupled satellite system. The location of this equilibrium does not need to coincide with the isolated-satellite prediction.

\subsubsection{Resonance Crossing}

Not all encounters with Jovian tidal resonances result in sustained resonance locking. A satellite may instead pass through the region of maximum tidal dissipation, entering and leaving on opposite sides of the resonance. During such a crossing, the migration timescale decreases sharply as the forcing frequency approaches the mode frequency, then increases after the satellite passes through resonance.

An example occurs in the Sharp-core simulation (Fig.~\ref{fig:smooth_sharp}b), where Io crosses two Jovian tidal resonances after Europa has become resonance locked. These encounters produce temporary changes in Io's migration rate and excite Europa's eccentricity, perturbing the libration of the Laplace angle. Europa remains locked to its Jovian mode during these encounters, and the Laplace resonance survives. This differs from temporary resonance locking, in which a satellite tracks the moving mode resonance for a finite interval before leaving from the same flank.

\section{Discussion}

\subsection{Dynamical Coupling and Eccentricity Excitation}

The migration of the resonant chain is driven by the combined tidal torques acting on the satellites, rather than the torque on the resonance-locked moon alone. Our angular-momentum budget calculations indicate that Europa can supply the majority of the external tidal angular-momentum input during its resonance-locking episodes, while Io's background tide provides an additional contribution. These torques are redistributed through the mean-motion resonances, allowing the satellites to migrate outward together.

These differences arise because the tidal torque on one moon also affects the orbits of the other two. When Io migrates outward more rapidly, it approaches Europa and can drive the resonant chain outward. A Ganymede lock has the opposite effect: Ganymede moves away from Europa, eventually disrupting their 2:1 resonance. As the middle moon, Europa's response to encountering a Jovian resonance is sensitive to the existing resonant configuration, and its outward migration can either be transmitted to both Io and Ganymede or leave Io behind, disrupting the inner resonance. 

The eccentricity attained during these interactions is not uniquely determined by which satellite is undergoing resonance locking. In our simulations, Io resonance locks generally produce greater eccentricity excitation than Europa locks, but the amplitude depends on the imposed tidal migration and eccentricity damping, as well as the preceding orbital evolution. 

Previous studies have shown that the dynamics of resonant systems can depend on the relative timescales of migration and eccentricity damping, which are parameterized by \(\tau_a/\tau_e\) (e.g., \citealt{bodman2014stability}). We report \(\tau_a/\tau_e\) as a useful diagnostic of the relative strength of these effects. However, our results show that the final eccentricity and libration state in our three-moon system are not determined uniquely by this ratio. Varying the $Q$ in the constant-$Q$ model simulations by a few orders of magnitude while keeping constant eccentricity-damping values can produce eccentricities of similar order. Further work is needed to characterize the structure of the Laplace resonance and understand how migration and eccentricity damping drive transitions across libration states. 

\subsection{Tectonic Implications for Europa}

The orbital histories produced by Jupiter's evolving tidal resonances imply time-dependent tidal heating inside Europa. In the representative simulations shown in Figures~\ref{fig:231_core0_params} and~\ref{fig:smooth_sharp}, inferred tidal power generally remains of order $10^{11}$--$10^{12}$ W, with shorter episodes approaching $\sim10^{13}$ W. These variations arise from changes in Europa's eccentricity as the satellites encounter and become locked to Jovian oscillation modes. The heating rates are calculated using Equation~\ref{eqn:edot} and our adopted eccentricity-damping timescale, rather than a coupled thermal-evolution model. For comparison, radiogenic heating in a chondritic Europa mantle is estimated to be $\sim2\times10^{11}$ W \citep{cammarano2006long}.

Resonance locking can maintain an elevated eccentricity for extended periods, whereas encounters with additional Jovian resonances can produce more abrupt heating excursions. Europa's thermal response would depend on the duration of these variations and on the dissipation and heat-transport properties of its interior. Thermal models have shown that changes in eccentricity can influence ice-shell thickness \citep{walker2022tidal}, although a self-consistent calculation would be needed to determine how Europa's shell responds to the orbital histories found here.

Changes in ice-shell temperature and thickness could also affect the stresses associated with cooling and refreezing \citep{nimmo2004stresses,rudolph2022cooling}. Europa's bands provide evidence of extensional deformation \citep{howell2018band,2014NatGe...7..762K}, although their formation cannot be attributed to specific heating episodes in our simulations.

Europa's surface preserves a record of tectonic activity that can be investigated through the morphology and cross-cutting relationships of bands and other lineaments \citep{haslebacher2014lineamapper}. The Europa Clipper's Europa Imaging System (EIS) will provide imaging and stereo topography to further constrain these relationships \citep{turtle2024eis}, while REASON will investigate the ice-shell structure and search for subsurface liquid water \citep{2024SSRv..220...51B}. Together, these observations may help constrain Europa's tectonic and thermal history. Connecting them to the orbital histories explored here will require models coupling tidal dissipation to Europa's thermal and mechanical evolution.

\subsection{Implications for Jupiter's Core}

The distribution of resonances depends sensitively on the assumed interior structure of Jupiter, which remains poorly constrained. We test three Jupiter core models: Compact, Smooth, and Sharp. The different tidal landscapes of these models produce sequences of resonance crossings and locks. The long-term survival of the 4:2:1 resonant chain may therefore provide an additional dynamical constraint on models of Jupiter's interior; however, our present calculations are not sufficient to use the Laplace resonance to exclude specific interior structures.

\subsubsection{Jupiter Model Limitations}

Our Jupiter models use first-order perturbation theory to incorporate the effects of planet rotation on mode frequency. This approach neglects the mode coupling caused by centrifugal oblateness and the Coriolis effect, introducing an important bias to the predictions of tidal resonance locations. We therefore avoid drawing conclusions that depend on the precise locations of individual Jovian tidal resonances.

Our calculation also does not capture the full spectrum of inertial and gravito-inertial modes. Inertial waves occupy the rotating-frame frequency range $|\omega|<2\Omega_J$, and their coupling to tidal forcing may introduce additional resonances or modify those already present. Coupling between different modes can also produce avoided crossings as Jupiter's interior evolves, potentially changing the resonance frequencies and their migration histories. More complete rotating planet models are needed to assess these effects.

\subsection{Model Limitations and Future Directions}

Our frequency-dependent tidal prescription assumes that Jupiter's tidal response follows the instantaneous forcing frequency. We evaluated the precomputed tidal migration timescale using each satellite's osculating mean motion, without explicitly evolving the amplitudes and phases of Jupiter's oscillation modes. A finite mode response time could alter the tidal torque during rapid passages through narrow resonances, which may affect the resulting orbital evolution. Future calculations could incorporate the time-dependent excitation and damping of individual Jovian modes rather than using a steady-state tidal profile.

We also do not model thermal feedback in the moons. In our frequency-dependent $Q$ integrations, eccentricity-damping timescales are held constant. The temporary increase in Io's damping timescale introduced in Section~4 is a prescribed, smooth variation intended to test the dynamical response, rather than a model of a specific thermal transition. In reality, enhanced tidal heating can change the thermal and rheological properties of the satellites, thereby modifying their dissipation \citep{peale1978contribution,ojakangas1986episodic, 2004Icar..171..391H}. Such feedback could alter the eccentricity excitation and duration of resonance-locking episodes. Coupled orbital and interior-evolution models will be needed to determine how these effects influence Europa's heating history \citep{bagheri2026jointdynamical}.

\section{Conclusions}

We developed a simple model that couples Jupiter's interior evolution to the dynamical evolution of the Galilean satellites. Encounters with Jovian tidal resonances can accelerate satellite migration, excite orbital eccentricity, and change the libration state or stability of the Laplace resonance.

The outcome depends on which satellite experiences resonance locking and the sequence of subsequent encounters. Locks involving Io or Europa can carry the resonant chain outward, while Ganymede locking destroys the Laplace resonance in the simulations considered here. Successive encounters can also replace an existing Jovian mode lock or produce temporary eccentricity excitation through resonance crossings. The distribution of tidal resonances differs between Jupiter's interior models, affecting the sequence of encounters and the resulting orbital evolution of the satellites. Further work is needed to determine whether these differences can be used to constrain Jupiter's internal structure.

The resulting eccentricity variations produce time-dependent tidal heating in Europa. In our representative simulations, the inferred background tidal power inside Europa is generally of the order $10^{11}$--$10^{12}$ W, with episodes reaching up to $10^{13}$ W under our adopted damping prescription. More complete models of Jupiter's rotating mode spectrum and the satellites' thermal evolution are needed to determine how these orbital histories may translate into Europa's tectonic evolution.

\section{Acknowledgments}

JS and BI acknowledge support from the Europa Clipper mission through the Europa ICONS program during the initial stages of this project. BI acknowledges support from the Juno mission. Computational resources were provided by the lux supercomputer at UC Santa Cruz, funded by NSF MRI grant AST 1828315. JS thanks the Lamat Institute for training and support during the development of this project, as well as friends and family for their support and patience.

\bibliography{sample701}{}
\bibliographystyle{aasjournalv7}

\appendix

\setcounter{figure}{0}
\renewcommand{\thefigure}{A.\arabic{figure}}

\section{Construction of the Frequency-Dependent Jovian Tidal Model}
\label{app:jupiter_tides}

\subsection{Jupiter Interior Models and Normal-Mode Frequencies}
\label{app:modes}

We consider the three representative dilute-core structures shown in Fig.~\ref{fig:taus}b.

The mode frequencies returned by the nonrotating calculation are corrected to first order in Jupiter's rotation. For a mode $\alpha$ with azimuthal number $m$, the rotation-corrected frequency is \citep{lai2021jupiter}
\begin{equation}
    \omega_\alpha
    =
    \omega_{\alpha 0}
    -
    m\Omega C_\alpha,
    \label{eq:rotational_correction}
\end{equation}
where $\omega_{\alpha 0}$ is the nonrotating mode frequency and
\begin{equation}
    C_\alpha
    =
    \frac{4\pi}{M_J}
    \int_0^{R_J}
    \rho r^2
    \left(
        2\xi_r\xi_h+\xi_h^2
    \right)
    dr.
    \label{eq:Calpha}
\end{equation}
Here, $\rho$ is the density profile of Jupiter and $\xi_r$ and $\xi_h$ are the radial and horizontal components of the mode displacement eigenfunction.

We allow the rotation-corrected mode frequencies to evolve according to the prescription introduced in Eq.~\ref{eq:mode_evolution},
\begin{equation}
    \omega_\alpha(t)
    =
    \left(
        \omega_{\alpha 0}
        -
        m\Omega C_\alpha
    \right)
    \exp\left(\frac{t}{t_{\rm evol}}\right).
    \label{eq:mode_evolution_full}
\end{equation}

Our treatment of rotation is approximate. Rotation can couple oscillation modes that would have distinct spherical-harmonic degrees in a nonrotating planet. Because we include rotation only through the first-order frequency correction in Eq.~\ref{eq:rotational_correction}, we neglect this mode mixing and retain only tidal forcing terms whose $(\ell,m)$ values match those assigned to each calculated mode. The consequences of this approximation for interpreting individual resonance locations are discussed in Section~6. 

\subsection{Tidal Excitation and Mode Damping}
\label{app:mode_damping}

For a satellite with mean motion $n$, the tidal forcing frequency in Jupiter's rotating frame is given by Eq.~\ref{eq:forcing_frequency}. A mode is resonantly excited when $\omega_f$ approaches its normal-mode frequency $\omega_\alpha$. The total energy-transfer rate is obtained by summing the contribution from each excited mode \citep{fuller2016resonance},
\begin{equation}
    \dot E
    =
    \sum_\alpha \dot E_\alpha,
    \label{eq:Edot_sum}
\end{equation}
where
\begin{equation}
    \dot E_\alpha
    =
    \frac{
        m R_J^2 \gamma_\alpha A_\alpha^2
    }{
        M_J
        \left[
            (\omega_\alpha-\omega_f)^2
            +
            \gamma_\alpha^2
        \right]
    }
    \left(
        \frac{n\omega_f}{\omega_\alpha^2}
    \right).
    \label{eq:Edot_mode}
\end{equation}

The tidal forcing amplitude of mode $\alpha$ is
\begin{equation}
    A_\alpha
    =
    \mathcal{G}m_s
    \left(
        \frac{R_J}{a}
    \right)^{\ell+1}
    Q_\alpha W_f,
    \label{eq:Aalpha}
\end{equation}
where $m_s$ is the mass of the perturbing satellite, $Q_\alpha$ measures the overlap between the oscillation mode and the tidal forcing, and $W_f$ is the geometric coefficient of the forcing potential.

The mode overlap is calculated as \citep{lai2021jupiter}
\begin{equation}
    Q_\alpha
    =
    4\pi
    \int_0^{R_J}
    r^{\ell+2}
    \delta\rho_\alpha^*
    dr,
    \label{eq:Qalpha}
\end{equation}
where $\delta\rho_\alpha$ is the Eulerian density perturbation associated with mode $\alpha$.

The angular dependence of the tidal forcing is described by \citep{lai2006resonant}
\begin{equation}
\begin{split}
    W_f
    =
    2^{-\ell}
    \left[
        \frac{
            4\pi(\ell+m)!(\ell-m)!
        }{
            2\ell+1
        }
    \right]^{1/2}
    \\
    \times
    \left[
        \left(
            \frac{\ell+m}{2}
        \right)!
        \left(
            \frac{\ell-m}{2}
        \right)!
    \right]^{-1},
\end{split}
\label{eq:Wf}
\end{equation}
for even $\ell+m$, and $W_f=0$ for odd $\ell+m$. In this work, we include forcing for $(\ell,m)=(2,2)$, $(4,4)$, $(4,2)$, $(3,3)$, and $(3,1)$.

We estimate the damping of each mode by turbulent convection using mixing-length theory, with a correction for the relationship between the tidal forcing and convective timescales \citep{goldreich1977turbulent,fuller2016resonance}. The damping rate is
\begin{equation}
    \gamma_\alpha
    =
    \frac{
        4\pi^2\nu_{\rm con}
    }{
        \omega_f t_{\rm conv}
    }
    f\left[
        \rho_\alpha r^2
        \left(
            \xi_r^2
            +
            \ell(\ell+1)\xi_h^2
        \right)
    \right],
    \label{eq:gammaalpha}
\end{equation}
where $\nu_{\rm con}$ is the convective viscosity and $t_{\rm conv}$ is the convective timescale. The functional $f$ weights the contribution from the convective portions of the planet,
\begin{equation}
\begin{split}
    f(x)
    =
    \left[
        \int_0^{R_{\rm bot}}
        k_r^2 x\,dr
        +
        \int_{R_{\rm top}}^{R_J}
        k_r^2 x\,dr
    \right]
    \left[
        \int_0^{R_J}
        x\,dr
    \right]^{-1},
\end{split}
\label{eq:functional_f}
\end{equation}
where $R_{\rm bot}$ and $R_{\rm top}$ define the boundaries of the convective regions. The radial wavenumber is approximated as
\begin{equation}
    k_r^2
    =
    \left(
        \frac{\omega_\alpha}{c_s}
    \right)^2
    -
    \frac{\ell(\ell+1)}{r^2},
    \label{eq:kr}
\end{equation}
where $c_s$ is the sound speed.

\subsection{Conversion to Orbital Migration}
\label{app:migration_profiles}

For a given satellite, we evaluate Eq.~\ref{eq:Edot_mode} for each included Jovian mode and sum the mode contributions according to Eq.~\ref{eq:Edot_sum}. The orbital migration timescale is then
\begin{equation}
    \tau_a
    =
    -\frac{E_{\rm orb}}
    {\sum_\alpha\dot E_\alpha},
\end{equation}
as given in Eq.~\ref{eq:mode_taua}. Evaluating this quantity as a function of satellite semi-major axis produces the initial $\tau_{a,0}(a)$ profiles shown in Fig.~\ref{fig:taus}a.

These profiles are calculated once for each Jupiter interior model. During the $N$-body integrations, we do not recompute Jupiter's oscillation modes. Instead, the precomputed tidal landscape is allowed to drift as the mode frequencies evolve according to Eq.~\ref{eq:mode_evolution}. The procedure used to map the evolving satellite orbit onto the precomputed $\tau_{a,0}(a)$ profile is described in Section~3.

\end{document}